\documentclass[titlepage,aps,showpacs,pra]{revtex4}

\usepackage{amsmath}
\usepackage{bm,amsfonts,graphicx}
\usepackage{amssymb}
\usepackage{amsmath}
\usepackage{graphicx}
\usepackage{color}
\usepackage{epstopdf}

\fontsize{10}{13}
\selectfont

\begin{document}
\fontsize{10}{13}
\selectfont
\title{Cross-Phase Modulation and Population Redistribution in a Periodic Tripod Medium}

\author{K. \surname{S\l owik}}
\email{karolina@fizyka.umk.pl}
\author{A. Raczy\'nski} \author{J. Zaremba}
\affiliation{Faculty of Physics, Astronomy and Informatics,
Nicolaus Copernicus University,
Toru\'n, Poland,}
\author{S. Zieli\'nska-Kaniasty}
\affiliation{Institute of Mathematics and Physics,
University of Technology and Life Sciences,
Bydgoszcz, Poland,}
\author{M. Artoni}
\affiliation{ Department of Physics and Chemistry of Materials
CNR-IDASC Sensor Lab, Brescia University, and European Laboratory
for Nonlinear Spectroscopy, Sesto Fiorentino, Italy. }
\author{G. C. La Rocca}
\affiliation{Scuola Normale Superiore and CNISM, Pisa, Italy.}

\begin{abstract}
The cross-Kerr effect is studied for two weak beams, probe and
trigger, propagating in an atomic medium in a tripod
configuration, dressed by a strong standing wave coupling beam in
a regime of electromagnetically induced transparency. The
nonlinear phase shifts for both transmitted and reflected probe
beams induced by the trigger's presence are found to depend on the
probe detuning, the control beams intensity, the relaxation rates
and, in particular, on the redistribution of the population among
the atomic levels.
 Such a quantitative analysis  indicates that the transmitted
 and reflected probe beam components and their respective phase shifts can be easily controlled and optimized.
\end{abstract}

\maketitle

\section{Introduction}

All-optical nonlinearities at low light levels are instrumental to
the optical implementation of quantum information processing
systems. Large nonlinearities can be achieved when the
light-matter interaction is resonant, in which case however
absorption typically plays a detrimental role. A great deal of
attention has thus been devoted to a variety of schemes based on
the regime of electromagnetically induced transparency (EIT)
\cite{fleischhauer} in which this counterbalance can be overcome.
In a typical EIT configuration, a strong control field, coupling
two unpopulated levels of a $\Lambda$ system, creates a
transparency window for a weak probe beam. This basic scheme was
extended by including additional atomic levels coupled by laser or
microwave fields in the double-$\Lambda$ \cite{my04}, tripod
\cite{paspalakis,mazets,my07}, $N$ \cite{kang,wilson} or
inverted-$Y$ \cite{joshi} configurations, just to name a few.
Under EIT-like conditions the weak interaction between two photons
(or weak classical pulses) may become largely enhanced. In
particular, very large cross nonlinear effects have been predicted
leading to new types of polarization phase gates in an
optically dressed medium in the $M$ \cite{ottaviani}, tripod
\cite{rebic, petr} or inverted-$Y$ \cite{joshi2} configuration.
Relevant experimental work has very recently been reported in Ref.
\cite{yang}. In these instances, a significant change of the phase
of one of the propagating pulses is achieved due to the cross-Kerr
effect induced by the other pulse. Analogous effects have recently
been studied, both theoretically and experimentally, for an
inverted-$Y$ system \cite{kou}, as well as for a four-level
$N$-type \cite{lo} and five-level system \cite{wang}. Important
developments concern also the dynamic control of the process
leading, e.g., to light slowdown, storage and release
\cite{liu,matsko}. These results are expected to pave the way
towards constructing all-optical logical devices.

Unlike most typical EIT configurations employing a running wave
coupling field, using a control standing wave coupling beam opens
the possibility of creating an all-optically tunable Bragg mirror.
In such a novel kind of dynamically tunable metamaterial
\cite{andre,my09} an incoming probe light encounters a spatially
periodic optical structure and is subject to Bragg scattering. For
specific frequency ranges transmission and stop bands appear, the
properties of which can be steered by a proper choice of the
control field. The field-induced dispersion of the medium and the
transmission and reflection spectra have been analyzed, e.g., in
Refs. \cite{andre,my09,myNowa}, including the case of a
quasi-standing wave coupling field \cite{artoni2}. It was also
possible to stop and store a pulse inside the medium, and then
retrieve it in the form of a stationary light pulse which could not
leave the medium due to the standing wave character of the
releasing control beam \cite{andre,bajcsy}. It has also been shown
that high nonlinearities can be achieved for stored light pulses
\cite{newchen}. An all-optical dynamic cavity for the confinement
of light pulses based on standing wave EIT Bragg mirrors  has
recently been proposed \cite{wucav}.

\vspace{2mm}

Very recently, we have pointed out that in the presence of a
trigger beam in a tripod configuration, a probe partially
reflected from the periodic structure induced by a strong standing
wave EIT coupling field  undergoes large cross-Kerr nonlinear
phase shifts. Such a novel configuration is specifically apt for
developing a phase-tunable beam splitter in which both the
amplitudes and the phases of the trasmitted and reflected beam can
be controlled. We have given some illustrative results valid in a
narrow range of frequencies owing to the restrictive assumption
that the atomic population is symmetrically distributed between
the two lower levels that are coupled with the upper level
respectively by the probe and trigger pulses \cite{myScripta}.
Here, we provide a comprehensive theoretical study of the problem,
including a proper treatment of the population redistribution. The
latter point turns out to be crucial to extend the range of
possible probe detunings and, thus, optimize the control
possibilities. Our approach combines analytical methodologies used
to describe propagation effects in both nonlinear and spatially
periodic media, and allow us to numerically examine the phase
shifts of the reflected and transmitted probe field induced by the
trigger's presence over a wide range of probe detunings as well as
their dependence on various parameters characterizing both the
atomic medium and the three laser fields. Our results indicate
that the tripod standing wave EIT configuration makes control over
the cross-Kerr effect more versatile than for a running wave EIT
configuration \cite{rebic, petr} in which no reflected beam
appears.

\vspace{2mm}

The paper is organized as follows.  In Section II, we present the
basic theoretical approach to calculate the cross-Kerr effect in a
medium additionally dressed by a quasi-standing wave EIT coupling
field, based on expanding the medium susceptibility into a power
series with respect to the probe and trigger fields and into a
Fourier series appropriate to the spatial periodicity imposed by
the coupling field. In Appendix A we derive in detail the formulae
for the susceptibility's expansion coefficients, while in Appendix
B we present the way to evaluate the population redistribution. In Section III,
we present and discuss the numerical results for the trigger
induced nonlinear phase shifts of the reflected and transmited
probe fields as functions of the probe detuning for various
combinations of control field intensities, relaxation rates and
sample length. Finally, we draw our conclusions.

\section{Theory}
\begin{figure}
\begin{center}
\includegraphics[scale=0.3]{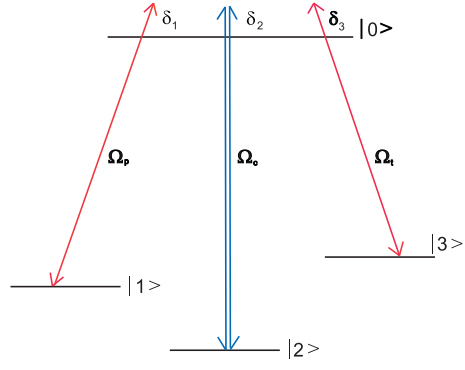}
\caption {\label{tri} (color online). The level and coupling scheme of the tripod configuration.}
\end{center}
\end{figure}

We consider a tripod system driven by a strong control field ($\mathcal{E}_c$) coupling the ground level $2$ and the upper level $0$, and by two weaker laser fields called probe ($\mathcal{E}_p$) and trigger ($\mathcal{E}_t$), coupling $0$ with $1$ and $3$, respectively, as shown in Fig. \ref{tri}. We express the fields through their complex amplitudes $\epsilon_{p,c,t}$ slowly varying in time:
$$\mathcal{E}_{p,c,t} = \epsilon_{p,c,t} e^{-i\omega_{1,2,3}t} + \epsilon_{p,c,t}^* e^{i\omega_{1,2,3}t}.$$
Each of the couplings is detuned by $\delta_j \equiv E_j+\hbar\omega_j-E_0$, $j = 1,2,3$. Here $E_{j}$ stands for the energy of the state $j$ and $\omega_j$ denotes the frequency of the $j$-th field.

All the fields propagate along the $z$ axis. In general the
control wave is allowed to have two counter-propagating
components of arbitrary ratio. In the case of a running control
wave, the component antiparallel to the incoming probe and trigger
fields is zero. In the case of a perfect standing wave, both
components are equal. Thus, the control field provides a spatial
lattice of period $\frac{\pi}{k_2}$, $k_2$ being the field's wave
number. In the configuration proposed below $k_2$ may be slightly
different from the probe and trigger fields' wave vectors
$k_{1,3}$, but that can be corrected by tilting both components of
the control field by an angle $\theta$ with respect to the
$z$-axis so that $\omega_{2}\cos\theta/c\approx k_{1,3}$
\cite{artoni2}.

Such a configuration can be realized in a cold gas of $^{87}$Rb atoms. The
states $1$ and $3$ may correspond, e.g., to Zeeman sublevels
$|5S_{\frac{1}{2}},F=2,m=\{-1,1\}>$, state $2$ to
$|5S_{\frac{1}{2}},F=1,m=1>$, and the upper state $0$ to the level
$|5P_{\frac{3}{2}},F=1,m=0>$. The scheme could be used as a
polarization phase gate, as was first proposed in Ref.
\cite{rebic}.  Each of the two pulses interacts with the medium
when it has the proper circular polarization: right for the probe
and left for the trigger. When both pulses are properly polarized
the nonlinear interaction between them gives rise to a cross-phase
modulation.

The Bloch equations for the atom + field system in the rotating wave approximation read:
\begin{eqnarray}
i\dot{\sigma}_{00} &=& \Omega_p \sigma_{10} +\Omega_c \sigma_{20} + \Omega_t \sigma_{30} - \Omega_p^* \sigma_{01} -\Omega_c^* \sigma_{02} - \nonumber \\
&-& \Omega_t^* \sigma_{03} -i(\gamma_{11}+\gamma_{22}+\gamma_{33})\sigma_{00}\nonumber \\
i\dot{\sigma}_{11} &=& \Omega_p^* \sigma_{01} - \Omega_p \sigma_{10} +i\gamma_{11}\sigma_{00} -i\gamma_{12}\sigma_{11} +i\gamma_{13}\sigma_{33}\nonumber \\
i\dot{\sigma}_{22} &=& \Omega_c^* \sigma_{02} - \Omega_c \sigma_{20} +i\gamma_{22}\sigma_{00} +i\gamma_{12}\sigma_{11} +i\gamma_{23}\sigma_{33}\nonumber \\
i\dot{\sigma}_{33} &=& \Omega_t^* \sigma_{03} - \Omega_t \sigma_{30} +i\gamma_{33}\sigma_{00} -i(\gamma_{13}+\gamma_{23})\sigma_{33} \nonumber \\
i\dot{\sigma}_{10} &=& \Delta_{10}^*\sigma_{10} - \Omega_p^*(\sigma_{11}-\sigma_{00}) -\Omega_c^* \sigma_{12} -\Omega_t^* \sigma_{13} \nonumber \\
i\dot{\sigma}_{20} &=& \Delta_{20}^*\sigma_{20} - \Omega_c^*(\sigma_{22}-\sigma_{00}) -\Omega_p^* \sigma_{21} -\Omega_t^* \sigma_{23} \label{beqs}\\
i\dot{\sigma}_{30} &=& \Delta_{30}^*\sigma_{30} - \Omega_t^*(\sigma_{33}-\sigma_{00}) -\Omega_c^* \sigma_{32} -\Omega_p^* \sigma_{31} \nonumber \\
i\dot{\sigma}_{12} &=& \Delta_{12}^*\sigma_{12} + \Omega_p^* \sigma_{02} - \Omega_c \sigma_{10} \nonumber \\
i\dot{\sigma}_{13} &=& \Delta_{13}^*\sigma_{13} + \Omega_p^* \sigma_{03} - \Omega_t \sigma_{10} \nonumber \\
i\dot{\sigma}_{23} &=& \Delta_{23}^*\sigma_{23} + \Omega_c^* \sigma_{03} - \Omega_t \sigma_{20} \nonumber
\end{eqnarray}
where $\sigma_{mm} = \varrho_{mm}$, $\sigma_{0j} = \varrho_{0j}e^{i\omega_jt}$, $\sigma_{jk} = \varrho_{jk}e^{-i(\omega_j-\omega_k)t}$, and $\varrho$ is the density matrix of the atoms in the Schr\"odinger picture, $\Delta_{j0} = \delta_j+i\gamma_{j0}$, $\Delta_{jk} = \delta_j-\delta_k +i\gamma_{jk}$, $\gamma_{jk} = \gamma_{kj}$, $j,k = 1,2,3$, $m = 0,1,2,3$. The Rabi frequencies are defined by $\Omega_{p,c,t} = - \frac{d_{01,02,03} \epsilon_{p,c,t}}{\hbar}$, where $d_{0j}$ is the electric dipole moment matrix element.

The simplified model of relaxations adopted in the above equations takes into account spontaneous emission from the upper state $0$ to a lower state $k$, described by the relaxation rates $\gamma_{0k}$ and $\gamma_{kk}$ with $\gamma_{0k} = \frac{1}{2} (\gamma_{11}+\gamma_{22}+\gamma_{33})$, as well as deexcitation and decoherence between the lower levels $\gamma_{jk}$, $j \neq k = 1,2,3$, due to atomic collisions. The relaxation rates for the coherences $\sigma_{jk}$, $j \neq k$, and those for the collision-induced population relaxations were taken equal for simplicity (cf. Ref. \cite{rebic}).

The propagation equation for the positive frequency part of the probe field, slowly varying in time, i.e. when $|\frac{\partial \epsilon_p}{\partial t}| << \omega_1|\epsilon_p|$, reads in the Fourier picture, with the Fourier frequency variable identified with the probe detuning
\begin{equation}
\left[ \frac{\partial^2}{\partial z^2}+\frac{1}{c^2} \left(\omega_1^2+2\omega_1\delta_1 \right)\right] \epsilon_p(z,\delta_1) = -\frac{\omega_1^2}{c^2}\chi_p(z,\delta_1) \epsilon_p(z,\delta_1), \label{tr} \end{equation}
with the medium susceptibility for the probe given by
\begin{equation}
\chi_p = -\lim_{t\rightarrow \infty} \frac{N|d_{10}|^2}{\hbar\epsilon_0} \frac{\sigma_{01}}{\Omega_p},
\end{equation}
where $N$ is the number of atoms per unit volume. The time limit means that we need to find the steady state solutions to the Bloch equations.

The assumption that the probe and trigger are weak with respect to the control field $|\Omega_p|^2, |\Omega_t|^2 << |\Omega_c|^2$ allows us to expand the susceptibility into the Taylor series:
\begin{equation}
\chi_p \approx \chi_p^{(1)} + \chi_{pp}^{(3)}|\Omega_p|^2 + \chi_{pt}^{(3)}|\Omega_t|^2. \label{taylor}
\end{equation}
The explicit form of the Taylor expansion coefficients can be found in Appendix A. The term $\chi_p^{(1)}$ corresponds to the linear susceptibility, while the third-order terms $\chi_{pp}^{(3)}$ and $\chi_{pt}^{(3)}$ represent the self- and cross-Kerr effect, respectively. The latter is of particular importance, as we are interested in the cross-phase modulation.

We now write the control field as:
\begin{equation}
\Omega_c = \Omega_c^+ e^{ik_2z} + \Omega_c^- e^{-ik_2z},
\end{equation}
where $\Omega_c^{\pm}$ of constant values are its forward and backward propagating parts. As the field is periodic in space, so are the medium optical properties. Therefore we expand the susceptibilities for both probe and trigger into the Fourier series:
\begin{equation}
\chi_{p} = \sum_{n=-\infty}^\infty\left[ \chi_{p,2n}^{(1)}+\chi_{pt,2n}^{(3)}|\Omega_t|^2+\chi_{pp,2n}^{(3)}|\Omega_p|^2\right] e^{2ink_2z}. \label{chip}
\end{equation}
The pulses propagating in a periodic medium acquire their reflected components. In the two-mode approximation (for details see \cite{artoni2, my09}) we write:
\begin{equation}
\Omega_{p,t} = \Omega^{+}_{p,t} e^{ik_2z} + \Omega^{-}_{p,t} e^{-ik_2z},\label{2m}
\end{equation}
where $\Omega^{\pm}_{p,t}$ are slowly varying in space.

We now insert Eqs. (\ref{chip},\ref{2m}) into Eq. (\ref{tr}) and drop terms rapidly oscillating in space. To write the propagation equations in the final form we make use of the relation $k_2 = \frac{\omega_2}{c}$ and set $\omega_2 = \omega_1+\Delta\omega_1$. For the probe we find:
\begin{eqnarray}
\left( i\frac{\partial}{\partial z} +\frac{\delta_1}{c} - \frac{\Delta\omega_1}{c} \right) \Omega_p^{+} &=& -\frac{\omega_1}{2c} \left(X\Omega_p^{+}+ Y\Omega_{p}^{-}\right), \label{epsplus} \\
\left( -i\frac{\partial}{\partial z} +\frac{\delta_1}{c} - \frac{\Delta\omega_1}{c} \right) \Omega_p^{-} &=& -\frac{\omega_1}{2c}\left(X\Omega_p^{-}+Z\Omega_{p}^{+}\right), \label{epsminus}
\end{eqnarray}
where
\begin{eqnarray}
X &=& \chi_{p,0}^{(1)}+\chi_{pt,0}^{(3)}S_t^2 +\chi_{pp,0}^{(3)}S_p^2 + \nonumber \\
&+& 2\chi_{pt,2}^{(3)}\Re(\Omega_t^{+}{\Omega_t^{-*}})+ 2\chi_{pp,2}^{(3)}\Re(\Omega_p^{+}{\Omega_p^{-*}}), \nonumber \\
Y &=& \chi_{p,2}^{(1)}+\chi_{pt,2}^{(3)}S_t^2 +\chi_{pp,2}^{(3)}S_p^2+\chi_{pt,0}^{(3)}\Omega_t^{+}{\Omega_t^{-*}}+ \nonumber \\
&+& \chi_{pt,4}^{(3)}\Omega_t^{-}{\Omega_t^{+*}}+\chi_{pp,0}^{(3)}\Omega_p^{+}{\Omega_p^{-*}}+\chi_{pp,4}^{(3)}\Omega_p^{-}{\Omega_p^{+*}}, \nonumber \\
Z &=& \chi_{p,2}^{(1)}+\chi_{pt,2}^{(3)}S_t^2+\chi_{pp,2}^{(3)}S_p^2+ \chi_{pt,0}^{(3)}\Omega_t^{+*}{\Omega_t^{-}}+\nonumber \\
&+&
\chi_{pt,4}^{(3)}\Omega_t^{-*}{\Omega_t^{+}}+\chi_{pp,0}^{(3)}\Omega_p^{+*}{\Omega_p^{-}}+\chi_{pp,4}^{(3)}\Omega_p^{-*}{\Omega_p^{+}},
\nonumber
\end{eqnarray}
with $S_{p,t}^2 \equiv |\Omega_{p,t}^+|^2 + |\Omega_{p,t}^-|^2$. The trigger equations are obtained by interchanging the indices $\{p,1\}\leftrightarrow\{t,3\}$. Explicit formulae for the susceptibilities Fourier components as well as their detailed deriviation can be found in Appendix A.

As we will change the probe detuning $\delta_1$ in a wide range
while keeping the trigger detuning $\delta_3$ constant, we
introduce an asymmetry in the populations $\sigma_{11}$ and
$\sigma_{33}$ even for $\Omega_p = \Omega_t$. Additionally, the
symmetry may be spoiled by the fact that $E_1$ (the energy of the
state $1$) is slightly smaller than $E_3$, and in a cold medium
transitions from $3$ to $1$ are possible while those from $1$ to
$3$ are not.  In previous works (see Refs. \cite{rebic,myScripta})
the probe and trigger fields were only allowed detunings such that
$\delta_1 \approx \delta_3$. In that case the stationary
population distribution was assumed to be $\sigma_{11} =
\sigma_{33} = \frac{1}{2}$. Such an approach is simple, but it is
well justified only in the resonance region of frequencies.  As we
show in the section \ref{results}, for a wider range of pulse
detunings it is necessary to make a better estimation of the
population distribution. In Appendix B we present the way to
evaluate the populations in the case of a quasi-standing coupling
beam ($\Omega_c^+\ne \Omega_c^-$, as in Ref.\cite{artoni2}).

\section{Results and discussion} \label{results}

In this section we present numerical results illustrating the cross-Kerr effect in our system and its dependence on the system parameters. In particular, we discuss how the cross-phase modulation is affected by the population redistribution.

We have solved the propagation equations (\ref{epsplus},\ref{epsminus}) with the boundary conditions corresponding to both probe and trigger originally propagating towards positive $z$:
\begin{eqnarray}
\Omega_{p,t}^+ (z=0,\delta_{1,3}) &=& {\Omega_0}_{p,t}, \label{bound}\\
\Omega_{p,t}^- (z=L,\delta_{1,3}) &=& 0, \nonumber
\end{eqnarray}
where ${\Omega_0}_{p,t}$ describes the amplitude of the incoming
probe/trigger pulse and $L$ is the sample length. We have used the susceptibilities'
Fourier components as presented in Appendix A. The level
populations have been evaluated as described in Appendix B, using
the initial amplitudes of the probe and trigger fields as input
data. Note that, strictly speaking, the populations in the
standing wave case are also modulated in space. However, as
discussed below, the latter effect should not play a very
important role in the case of an quasi-standing wave.

The equations are solved iteratively until self-consistency is achieved, alternately for the probe and trigger, with the susceptibilities calculated using the fields obtained in the preceding step. In the first iteration for the probe the initial constant value of the trigger has been used. The numerical task in each step is thus solving a linear equation. We find two independent solutions with one-point boundary conditions and combine them to obtain the solution with the proper two-point boundary conditions given by Eqs. (\ref{bound}).

The transmission and reflection coefficients for the probe and trigger beams are defined by:
\begin{eqnarray}
T_{p,t}(\delta_{1,3}) &=& \left| \frac{\Omega_{p,t}^+(L,\delta_{1,3})}{{\Omega_0}_{p,t}} \right|^2, \\
R_{p,t}(\delta_{1,3}) &=& \left| \frac{\Omega_{p,t}^-(0,\delta_{1,3})}{{\Omega_0}_{p,t}} \right|^2, \nonumber
\end{eqnarray}
while the phases of the transmitted and reflected fields are given by:
\begin{eqnarray}
\varphi^+_{p,t}(\delta_{1,3}) &=& \arg (\Omega_{p,t}^+(L,\delta_{1,3})), \\
\varphi^-_{p,t}(\delta_{1,3}) &=& \arg (\Omega_{p,t}^-(0,\delta_{1,3})). \nonumber
\end{eqnarray}

Unless otherwise specified, we use the following set of input data
which are of order of those of typical atomic systems: $\Omega_c^+
= 4$ MHz, $\Omega_c^- = 2$ MHz, $\Omega_{p0} = \Omega_{t0} = 0.67$
MHz, $\gamma_{10,20,30} = 0.67$ MHz, $\gamma_{12,32,13} = 6.67
\times 10^{-4}$ MHz, $\gamma_{11,22,33} = 0.44$ MHz, $\delta_2 =
6.67$ MHz, $\delta_3 = 1.002 \times 6.67$ MHz, $L = 1.06$ mm,
$d_{10} = d_{30} = 8 \times 10^{-30}$ C$^.$m, $N = 1.3 \times
10^{13}$ cm$^{-3}$. Our choosing a quasi-standing wave is
connected with the fact that for a perfect one the steady state
population in the node regions is trapped in the state $2$ due to
relaxation effects and the medium becomes transparent to both
probe and trigger. Therefore, it is reasonable to only consider
the cases in which the control field at the quasinodes is still
strong enough to pump the population from the level $2$.

\begin{figure}
\includegraphics[scale=0.9]{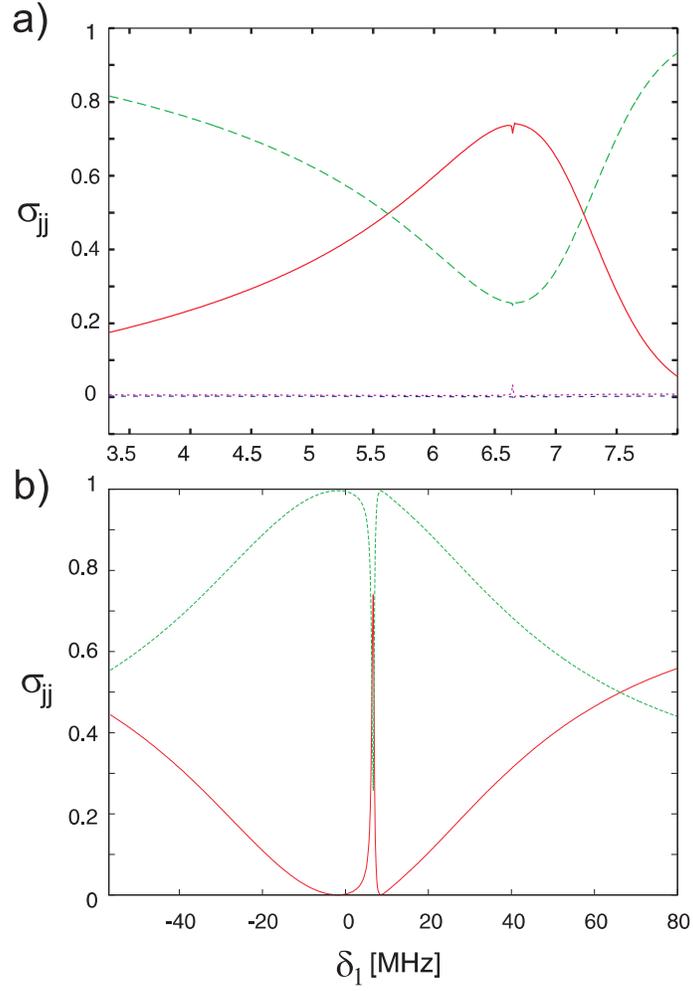}
\caption{\label{2} (color online). The population distribution in the case of trigger present: $\sigma_{11}$ - solid red line, $\sigma_{33}$ - dashed green line, $\sigma_{00}$ - short-dashed blue line, $\sigma_{22}$ - dotted violet line, for a narrow (a) and wide (b) frequency range. The populations $\sigma_{00}$ and $\sigma_{22}$ are negligible and thus not shown in the case (b).}
\end{figure}

We first calculate the corrected values of the populations which
are shown in Fig. \ref{2} in a narrow (plot (a)) and wide (plot
(b)) ranges of the probe detunings. Note that the populations
$\sigma_{11}$ and $\sigma_{33}$ may significantly differ from
$\frac{1}{2}$. They may vary rapidly within the transparency
window and the maximum value of $\sigma_{33}$ may be close to
unity which apparently occurs when the frequency of the probe
suits the energy interval between $E_0$ and the energy of one of
the lower states dressed by $\Omega_c$ and $\Omega_t$. Between the
maxima there is a deep minimum of $\sigma_{33}$ due to the
population trapping in the state $1$. The populations of
$\sigma_{00}$ and $\sigma_{22}$ are negligible, which is
consistent with our assumption.

If there are no relaxations in the system, then the steady state
solution for the no-trigger case is trivial: $\sigma_{33} = 1$,
$\sigma_{11} = \sigma_{00} = \sigma_{22} = 0$ due to spontaneous
emission from the upper level to all the lower levels. There is no
mechanism yet pumping the population out from $3$. The
nonnegligible relaxations provide such a mechanism and the plot
illustrating the $\sigma_{33}$ dependence on the probe detuning is
of a similar shape as in the trigger-present case (not shown).

\begin{figure}
\includegraphics[scale=0.8]{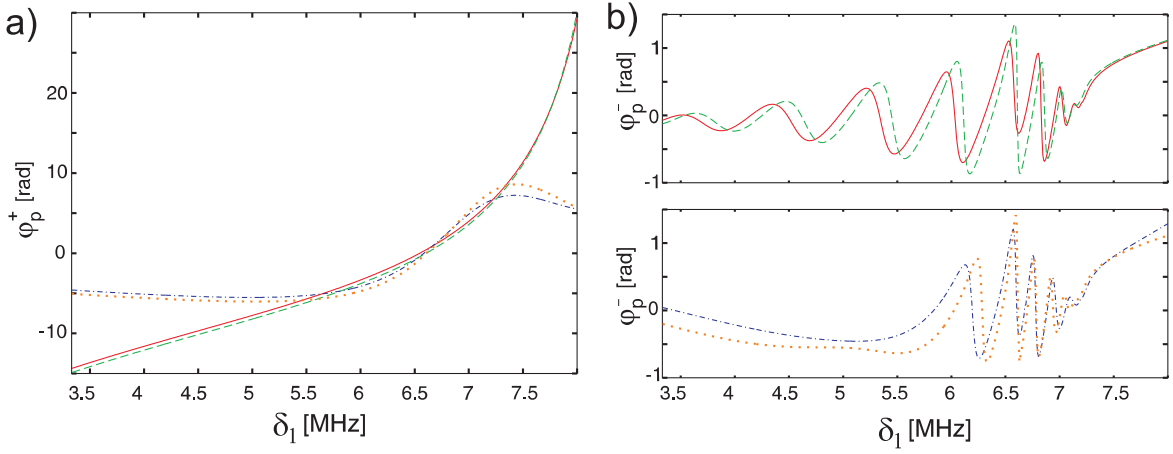}
\caption{\label{3} (color online). Probe field's phase shift
$\varphi_p(\delta_1)$ vs the probe detuning in the presence (solid
red) or absence (dashed green) of the trigger for balanced
population distributions $\sigma_{11} = \sigma_{33} = 0.5$. Phase
shift in the presence (dash-dotted blue) or absence (dotted
orange) of the trigger when instead the population distribution is
given by Eqs. (B.3). The frame (a) refers to the transmitted probe
phase shift ($\varphi_p^+(\delta_1)$) acquired in the running wave
configuration, whereas the frame (b) refers to the reflected probe
phase shift ($\varphi_p^-(\delta_1)$) acquired in the
quasi-standing wave configuration. }
\end{figure}

The phase shifts for a running control field ($\Omega_c^+ = 4$
MHz, $\Omega_c^- = 0$, the other parameters unchanged) are shown
in Fig. \ref{3}a. We compare the results for the transmitted probe
beam in the cases of equal or corrected values of the populations
$\sigma_{11}$ and $\sigma_{33}$ and for the trigger pulse present
or absent. The values of the obtained phase shifts
$\varphi_p^+(\delta_1)$ differ significantly for the two
approximations concerning the populations, except for three
points: when the calculated population distribution is almost
exact: $\sigma_{11} = \sigma_{33} \approx 0.5, \sigma_{22} =
\sigma_{00} \approx 0$ and within the resonance region $\delta_1
\approx \delta_2 \approx \delta_3$. The trigger-induced (Kerr)
phase shifts (the differences between the values given by pairs of
curves) in the two cases are also different and there are
frequency intervals in which the Kerr shifts are significantly
larger in the case in which the populations have been calculated
according to Eqs. (\ref{diag1}). In this way the results of Ref.
\cite{rebic} can be generalized for a wider range of the probe
detunings.

In Fig. \ref{3}b we make a similar comparison for a reflected beam
in the case of a quasi - standing control field ($\Omega_c^+ =
4.19$ MHz, $\Omega_c^- = 2$ MHz, the other parameters unchanged).
Again taking into account the population redistribution (lower
plot) caused a qualitative change of the behaviour of the phase
shifts as well as of their differences (the Kerr shifts) compared
with the case of equal populations. The results are in agreement
at the same three points as before. However, at $\delta_1 \approx
\delta_2 \approx \delta_3$, due to a numerical instability in the
algorithm used to find the solutions of Eq. (\ref{diag1}), the
calculated density matrix is no longer positive definite and our
results in this very narrow spectral region are unphysical (see
the sharp spikes appearing in Fig. \ref{2}).

\begin{figure}
\includegraphics[scale=0.9]{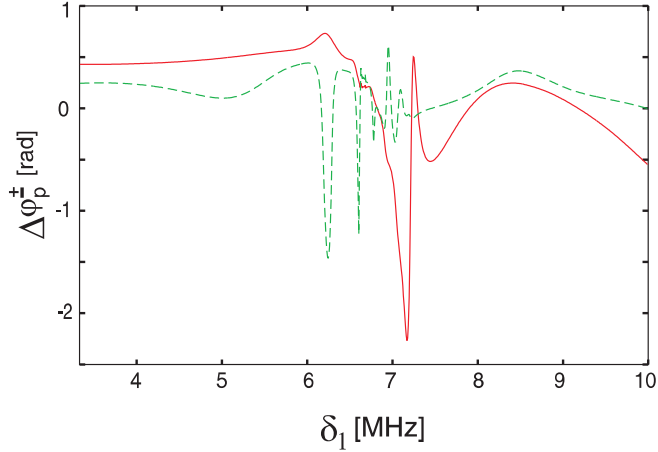}
\caption{\label{4} (color online). Trigger-induced probe phase
shifts $\Delta \varphi^\pm_p$ as a function of the probe detuning
for  transmission (solid red) and reflection (dashed green). The
population distribution is given by Eqs. (\ref{diag1}) while all
other parameters are as in Fig. \ref{3}b.}
\end{figure}

\begin{figure}
\includegraphics[scale=0.9]{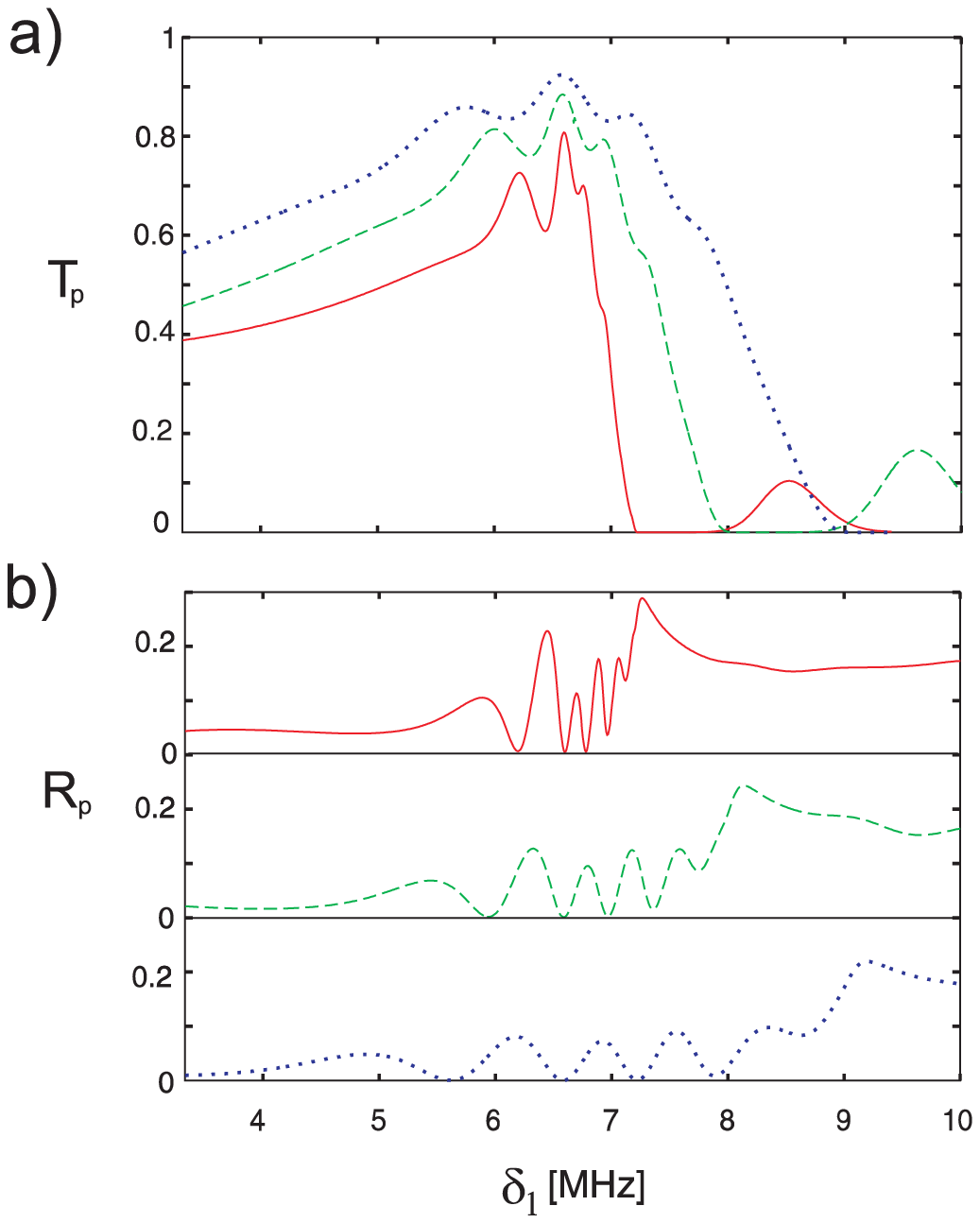}
\caption{\label{6} (color online). Transmission (a) and reflection
(b) coefficients in the presence of a trigger and for different
values of the control field $\Omega_c^+ = 4$ MHz  (solid red),
$\Omega_c^+ = 5.33$ MHz (dashed green), $\Omega_c^+ = 6.67$ MHz
(dotted blue) when the other coupling beam is set to $\Omega_c^- =
2$ MHz. }
\end{figure}

In Fig. \ref{4} we show the cross-Kerr phase shifts $\Delta
\varphi^\pm_p(\delta_1)$ due to the trigger's presence for both
the transmitted and reflected probe beams defined as
$\Delta\varphi^\pm_p=\varphi_p^\pm$(trigger
on)$-\varphi_p^\pm$(trigger off). One can see that the shifts may
be of order of one radian and there is a frequency range in which
they do not change rapidly. The narrow minima or maxima in Fig.
\ref{4} correspond to the situation in which the phases
$\varphi^\pm_p$ vary rapidly. The present results show how our
system behaves as an all-optically controlled Kerr medium. In
particular, it may serve as a tunable beam splitter for the probe
beam in which the phases of the reflected and/or trasmitted beam
are significantly affected by the trigger beam. For example, for
detunings of $\delta_1 \sim 5.5$ MHz or of $\delta_1 \sim 6.4$ MHz
at which both the transmission and reflection coefficients are
large (see Fig. \ref{6}), the trigger induced phase shift of the
trasmitted and reflected probe beams are both appreciable. On the
contrary, for a detuning of $\delta_1 \sim 6.25$ MHz at which both
the transmission and reflection coefficients are considerable (see Fig.
\ref{6}), while the trigger induced phase shift for the
transmitted probe beam is significant, that of the reflected beam
is negligible. This beam splitting action accompanied by a trigger
induced phase shift of the reflected and/or transmitted probe
components is a unique feature of the tripod standing wave EIT
configuration with respect to others previously considered. Taking
into account the redistribution of the population among the atomic
levels is however crucial. If the redistribution were ignored the
phase shifts could only be estimated within rather limited ranges
of frequencies. In particular, while for $\delta_1 \sim 5.5$ MHz
the approximation $\sigma_{11}=\sigma_{33}=0.5$ is still
reasonable as shown in Fig. \ref{2}, for  $\delta_1 \sim 6.4$ MHz
or $\delta_1 \sim 8.7$ MHz it fails. In the following, in order to
illustrate the dependence of the Kerr effect on various
parameters, we discuss several numerical results obtained over a
large detuning range using the method here developed to calculate
the population redistribution.

\begin{figure}
\includegraphics[scale=0.78]{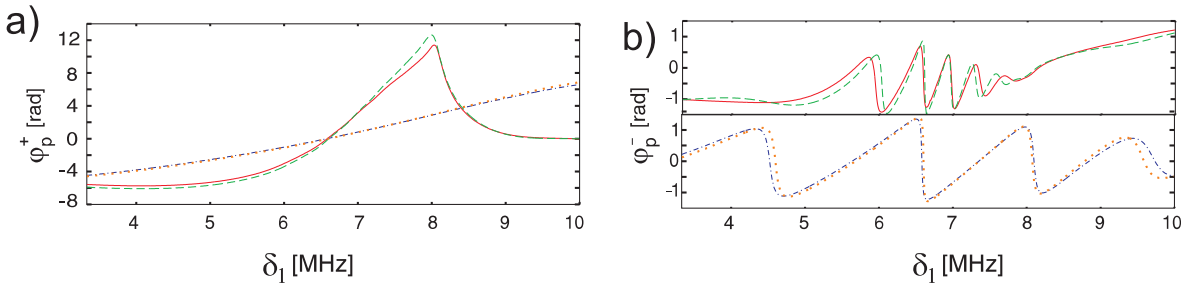}
\caption{\label{5} (color online). A comparison of the trigger-induced probe field's phase shifts (a) $\varphi_p^+(\delta_1)$, (b) $\varphi_p^-(\delta_1)$, vs the probe detuning $\delta_1$ for different control fields: $\Omega_c^+ = 6.67$ MHz, trigger present - solid red line, trigger absent - dashed green line, $\Omega_c^+ = 10$ MHz, trigger present - dot-dashed blue line, trigger absent - dotted orange line (cf. Fig. \ref{3}b, lower plot, in which $\Omega_c^+ = 4$ MHz).}
\end{figure}

We now check how the phase shifts of both the transmitted field
and reflected one depend on 'how much standing' the control field
is. We change the right-propagating part of the control field
$\Omega_c^+$ while keeping the left-moving part constant:
$\Omega_c^- = 2$ MHz. All the other parameters remain unchanged.
The results are shown in Fig. \ref{5}a for the transmitted beam
and Fig. \ref{5}b for the reflected one. All the plots cross in
the resonance region where the phase shift is small. The phase
shift $\varphi_p^+$ reaches a maximal value for some $\delta_1$
which depends on $\Omega_c^+$. The more intense is the control
field the flatter is the plot and the larger is the peak's shift
towards higher frequencies, which provides a way of controlling
the phases. We find the same effect of the peak moving to the
right when decreasing the left-propagating part of the control
field $\Omega_c^-$ while keeping $\Omega_c^+$ constant (not
shown). The more intense is the control field the smaller is the
trigger-induced phase shift. Around the resonance the reflected
phase shift $\varphi_p^-$ shows an oscillatory dependence on the
probe detuning. The width of the frequency range where the
oscillations are present grows with $\Omega_c^+$ but the number of
peaks remains constant - again the phase shift's plot becomes
flatter when the control field is increased. The curves are
complicated now but roughly we can say that the trigger-induced
phase shift decreases when the control field becomes stronger. The
extrema of the Kerr phase shifts $\Delta \varphi^-_p$ in the
intervals in which $\varphi_p^-$ vary rapidly become less
prominent and the distance between them increases for growing
control fields. In general a too strong control field is not
advantegous for generating considerable trigger-induced phase
shifts. Then the results gradually turn into those typical of the
usual EIT case.

As expected, the transmission coefficient (see Fig. \ref{6}a)
grows with $\Omega_c^+$ and so does the width of the transparency
window. Again the plots corresponding to the reflection are more
complex, as shown in Fig. \ref{6}b. In general the reflection
coefficient decreases when $\Omega_c^+$ is increased, but there
are some frequency ranges where the dependence is more
complicated. Yet if both parts of the control field are increased
simultaneously, so that $\frac{\Omega_c^-}{\Omega_c^+}$ is
constant, then not only the transmission but also the reflection
coefficient increases for most of the probe frequencies. This is
due to the absorption cancellation by a strong control field.

We have also checked how our results depend on the relaxation
rates $\gamma_{ij}$, $j \neq i =1,2,3$, due to interatomic
collisions. The main observation is that the trigger-induced phase
shift decreases for growing $\gamma_{ij}$. On the other hand, as
the relaxation rates grow, the frequency range where the reflected
field's phase dependence is oscillatory becomes wider and the
number of oscillations increases. As expected, the transmission
and reflection coefficients decrease in general when the
relaxation rates grow, but there are some frequency ranges where
the reflection coefficient is a nonmonotonic function of
$\gamma_{ij}$. Populations $\sigma_{00}$ and $\sigma_{22}$ are
negligible except for very large values of the relaxations
($\gamma_{ij} \sim 10^{-1}$ MHz) when they are of the order of a
few percent. For increasing $\gamma_{ij}$ the phase shifts become
less steep and an approximation consisting in adopting constant
values of the populations $\sigma_{11}$ and $\sigma_{33}$ may be
better justified.

Increasing the length of the sample leads to increasing the
trigger-induced phase shift of the transmitted field. For the
considered lengths of the sample the nonlinear phase shift
increases for growing $L$. However, for an extended sample both
pulses become partially absorbed and their mutual impact is
smaller. Elongating the sample leads to no significant change in
the amplitude of oscillations in the reflected field's phase but
it does influence the reflection coefficient in a nonmonotonic
way, which provides another way of controlling the cross-Kerr
effect. Thus, in order to have a considerable phase shift the
coupling field cannot be too strong and the relaxation rates
between the lower tripod states should not be too large; also the
length of the sample should be appropriately optimized.

\section{Conclusions}

We have presented a comprehensive study of the cross Kerr effect
in the propagation of two weak laser fields, a trigger and a
probe, in a medium of four-level atoms in the tripod configuration,
dressed by a third strong coupling field in the quasi-standing
wave configuration. This has been done by taking into the proper
account the population redistribution. Using the relevant terms of
the Fourier expansion of the linear and nonlinear
susceptibilities, we have numerically solved the propagation
equations for the right- (incoming) and left- (reflected) running
components of the two weak fields. In particular, the phase shifts
of the transmitted and reflected probe induced by the trigger turn
out to be as large as one radian. We have further shown that the
population redistribution significantly affects the phase shifts
of the transmitted and reflected beams unless very specific values
of probe detuning are used. Over a wide range of probe detunings,
on the contrary, we find that the trigger induced phase shifts for
the reflected and transmitted probe can be flexibly and
independently changed.
 We have finally characterized the
Kerr effect's dependence on the strength of the coupling field,
the relaxation rates and the sample's length.

In summary, by using a quasi-standing wave configuration the
amplitudes of the transmitted and reflected parts of the probe
beam and their respective trigger-induced phase shifts can be
controlled and optimized by acting on various parameters, and
especially on the probe detuning. While previous work has shown
the usefulness of the tripod configuration in a running wave EIT
regime to achieve a large cross-Kerr effect on the transmitted
probe beam, our results extend such findings to the standing wave
EIT regime in which both the reflected and transmitted probe beams
can be separately controlled.

\begin{acknowledgments}
K. S\l owik is grateful to G. C. La Rocca and Scuola Normale
Superiore in Pisa for their kind hospitality.  The work of K. S\l
owik was sponsored by the scholarship for doctoral students ZPORR
2008/2009 of the Marshal of the Kuyavian-Pomeranian Voivodeship.
Financial support from grant Azione integrata IT09L244H5 of MIUR
is also acknowledged.

\end{acknowledgments}
\vspace*{1cm}
\begin{center}{APPENDIX A} \end{center}
\setcounter{equation}{0}
\renewcommand{\theequation}{A.\arabic{equation}}
The medium susceptibilities can be written as:
\begin{eqnarray}
\chi_p &=& -\lim_{t\rightarrow \infty} \frac{N|d_{10}|^2}{\hbar\epsilon_0} \frac{\sigma_{01}}{\Omega_p}, \nonumber \\
\chi_t &=& -\lim_{t\rightarrow \infty} \frac{N|d_{30}|^2}{\hbar\epsilon_0} \frac{\sigma_{03}}{\Omega_t}. \nonumber
\end{eqnarray}
and can be calculated with the use of the steady state solutions of the Bloch equations.

From the last three of Eqs. (\ref{beqs}) we find the steady state spin coherences:
\begin{eqnarray}
\sigma_{12} = \frac{\Omega_c \sigma_{10} - \Omega_p^* \sigma_{02}}{\Delta_{12}^*}, \nonumber \\
\sigma_{32} = \frac{\Omega_c \sigma_{30} - \Omega_t^* \sigma_{02}}{\Delta_{32}^*}, \nonumber \\
\sigma_{13} = \frac{\Omega_t \sigma_{10} - \Omega_p^* \sigma_{03}}{\Delta_{13}^*}. \nonumber
\end{eqnarray}
We substitute these expressions into the stationary form of the equations $5-7$ in Eqs. (\ref{beqs}) and thus obtain a set of three coupled equations for $\sigma_{10}$, $\sigma_{20}$, $\sigma_{30}$. As the control field is much stronger than any other field present in the system, it prevents the states $0$ and $2$ from being populated. Therefore we set $\sigma_{00} = \sigma_{22} = 0$, as was done in Ref. \cite{rebic}. However, contrary to Ref. \cite{rebic}, here we do not assume the probe and trigger detunings to be almost equal. Hence the population distribution between the states $1$ and $3$ may be asymmetric and in general $\sigma_{11} \neq \sigma_{33}$. This crucial issue has been discussed in Section \ref{results}. Next we eliminate $\sigma_{20}$ and arrive at:
$$\left( \Delta_{10} -\frac{|\Omega_c|^2}{\Delta_{12}}-\frac{|\Omega_t|^2}{\Delta_{13}} + \frac{|\Omega_c|^2|\Omega_p|^2}{D\Delta_{12}^2}\right)\sigma_{01} +$$ $$+\frac{\Omega_t\Omega_p}{\Delta_{13}}\sigma_{30}+\frac{|\Omega_c|^2\Omega_p\Omega_t^*}{D\Delta_{12}\Delta_{32}}\sigma_{03} = \Omega_p \sigma_{11}$$
$$\left( \Delta_{30} -\frac{|\Omega_c|^2}{\Delta_{32}}-\frac{|\Omega_p|^2}{\Delta_{31}} + \frac{|\Omega_c|^2|\Omega_t|^2}{D\Delta_{32}^2}\right)\sigma_{03} +$$ $$+\frac{\Omega_t\Omega_p}{\Delta_{31}}\sigma_{10}+\frac{|\Omega_c|^2\Omega_t\Omega_p^*}{D\Delta_{12}\Delta_{32}}\sigma_{01} = \Omega_t \sigma_{33}$$
where $D = \Delta_{20}^*+\frac{|\Omega_p|^2}{\Delta_{12}}+\frac{|\Omega_t|^2}{\Delta_{32}}$.

We combine the above expressions to obtain $\frac{\sigma_{01}}{\Omega_p}$ and $\frac{\sigma_{03}}{\Omega_t}$. The solutions of the above set of equations expanded into the Taylor series have the form of Eq. (\ref{taylor}), with:
\begin{eqnarray}
\chi_p^{(1)}(\delta_1) &=& -\frac{N|d_{10}|^2\sigma_{11}}{\hbar\epsilon_0} \frac{\Delta_{12}}{\Delta_{10}\Delta_{12}-|\Omega_c|^2}, \label{chi1}\\
\chi_{pp}^{(3)}(\delta_1) &=& \frac{N|d_{10}|^2}{\hbar \epsilon_0} \frac{\sigma_{11}}{\Delta_{20}^*} \frac{|\Omega_c|^2}{(\Delta_{10}\Delta_{12}-|\Omega_c|^2)^2}, \\
\chi_{pt}^{(3)}(\delta_1) &=& -\frac{N|d_{10}|^2}{\hbar\epsilon_0}\left[ \frac{\Delta_{12}}{\Delta_{13}} \frac{1}{\Delta_{10}\Delta_{12}-|\Omega_c|^2} \right.\times \nonumber \\
&\times&\left( \frac{\sigma_{11}\Delta_{12}}{\Delta_{10}\Delta_{12}-|\Omega_c|^2} + \frac{\sigma_{33}\Delta_{23}}{-\Delta_{30}^*\Delta_{23}-|\Omega_c|^2}\right) - \label{chi3} \\
&-& \left.\frac{\sigma_{33}}{\Delta_{20}^*} \frac{|\Omega_c|^2}{(\Delta_{10}\Delta_{12}-|\Omega_c|^2)(\Delta_{30}\Delta_{32}-|\Omega_c|^2)}\right]. \nonumber
\end{eqnarray}
The trigger susceptibilities are obtained by interchanging the indices $\{p,1\}\leftrightarrow\{t,3\}$.

For the control field being in the form of a quasi-standing wave $\Omega_c = \Omega_c^+ e^{ik_2z} + \Omega_c^- e^{-ik_2z}$ the medium susceptibilities can be then expanded into the Fourier series:
\begin{equation}
\chi_{p}^{(1)} = \chi_{p,0}^{(1)}+2 \sum_{n=1}^{\infty} \chi_{p,2n}^{(1)} \cos(2nk_cz), \nonumber
\end{equation}
and similarly for $\chi_{pt,2n}^{(3)}$, etc. To find the explicit form of the Fourier coefficients we rewrite the expressions (\ref{chi1} - \ref{chi3}):
\begin{eqnarray}
\chi_{p}^{(1)} &=&  \frac{N|d_{10}|^2 \sigma_{11}}{\hbar \epsilon_0} \frac{A\Delta_{12}}{1+B\cos(2k_cz)}, \nonumber \\
\chi_{pt}^{(3)} &=& - \frac{N|d_{10}|^2}{\hbar \epsilon_0} \left[\frac{\nu}{(1+B \cos(2k_cz))^2} \right.- \nonumber \\
&-& \frac{\eta A}{1+B\cos(2k_cz)} + \frac{\eta C}{1+D\cos(2k_cz)}+ \nonumber \\
&+& \left(\frac{\eta' A}{1+B \cos(2k_cz)}-\frac{\eta' C^*}{1+D^*\cos(2k_cz)}\right)\times \nonumber \\
&\times& \left.\left(S_c^2+2\Omega_c^
+\Omega_c^-\cos(2k_cz)\right)\right], \nonumber \\
\chi_{pp}^{(3)} &=& \frac{N|d_{10}|^2}{\hbar \epsilon_0}\frac{\nu'}{\Delta_{20}^*} \frac{S_c^2+2\Omega_c^+\Omega_c^- \cos 2k_cz}{\left[1+B\cos(2k_cz)\right]^2}, \nonumber
\end{eqnarray}
where: $A=\left(S_c^2-\Delta_{10}\Delta_{12}\right)^{-1}$, $B= 2\Omega_c^+\Omega_c^-A$, $S_c^2 ={\Omega_c^+}^2+{\Omega_c^-}^2$, $\nu = \frac{\Delta_{12}^2}{\Delta_{13}}\nu'$, $\nu' = \sigma_{11} A^2$, $\eta = -\frac{\sigma_{33}\Delta_{12}\Delta_{23}/\Delta_{13}}{\Delta_{30}^*\Delta_{23}+\Delta_{10}\Delta_{12}}$, $\eta' = \frac{\sigma_{33}/\Delta_{20}^*}{\Delta_{30}\Delta_{32}-\Delta_{10}\Delta_{12}}$, $C = \left(S_c^2+\Delta_{30}^*\Delta_{23}\right)^{-1}, D = 2\Omega_c^+\Omega_c^-C$.

Let us denote:
\begin{eqnarray}
F^{(1)}_n(\mu) &\equiv& \frac{1}{2\pi} _{-\pi}\int^{\pi} \frac{\cos(nx)dx}{1+\mu\cos(x)} = \frac{1}{\sqrt{1-\mu^2}} \left( \frac{\sqrt{1-\mu^2}-1}{\mu}\right)^n,\nonumber \\
F^{(2)}_0(\mu) &\equiv& \frac{1}{2\pi} _{-\pi}\int^{\pi} \frac{dx}{[1+\mu\cos(x)]^2} = \frac{1}{\sqrt{1-\mu^2}^3}, \nonumber \\
F^{(2)}_1(\mu) &\equiv& \frac{1}{2\pi} _{-\pi}\int^{\pi} \frac{\cos(x)dx}{[1+\mu\cos(x)]^2} = \frac{-\mu}{\sqrt{1-\mu^2}^3}, \nonumber \\
F^{(2)}_2(\mu) &\equiv& \frac{1}{2\pi} _{-\pi}\int^{\pi} \frac{\cos(2x)dx}{[1+\mu\cos(x)]^2} = \frac{2}{\mu^2}+\frac{3-\frac{2}{\mu^2}}{\sqrt{1-\mu^2}^3}, \nonumber \\
F^{(2)}_3(\mu) &\equiv& \frac{1}{2\pi} _{-\pi}\int^{\pi} \frac{\cos(3x)dx}{[1+\mu\cos(x)]^2} = -\frac{8}{\mu^3} +\frac{\frac{8}{\mu^3}-\frac{12}{\mu}+3\mu}{\sqrt{1-\mu^2}^3}.\nonumber
\end{eqnarray}
The above integrals have been calculated with the use of the residues method. The Fourier coefficients of the susceptibilities read then:
\begin{eqnarray}
\chi_{p,2n}^{(1)} &=& \frac{N\sigma_{11}|d_{10}|^2}{\hbar \epsilon_0} A\Delta_{12} F^{(1)}_n(B), \nonumber \\
\chi_{pt,0}^{(3)} &=& - \frac{N|d_{10}|^2}{\hbar \epsilon_0} \left[ \nu F^{(2)}_0(B)\right.- \nonumber \\
&-&\eta \left(AF^{(1)}_0(B)-CF^{(1)}_0(D)\right)\nonumber \\
&+& \eta' S_c^2\left(AF^{(1)}_0(B)-C^*F^{(1)}_0(D^*)\right)+ \nonumber \\
&+&\left.\eta'2\Omega_c^+\Omega_c^- \left(AF^{(1)}_1(B)-C^*F^{(1)}_1(D^*) \right) \right], \nonumber \\
\chi_{pt,2}^{(3)} &=& - \frac{N|d_{10}|^2}{\hbar \epsilon_0} \left[ \nu F^{(2)}_1(B) \right.- \nonumber \\
&-& \eta \left(AF^{(1)}_1(B)-CF^{(1)}_1(D)\right)\nonumber \\
&+& \eta'S_c^2 \left(AF^{(1)}_1(B)-C^*F^{(1)}_1(D^*)\right)+ \nonumber \\
&+& \eta'\Omega_c^+\Omega_c^-A\left(F^{(1)}_0(B)+F^{(1)}_2(B)\right)- \nonumber \\
&-& \left.\eta'\Omega_c^+\Omega_c^-C^*\left(F^{(1)}_0(D^*)+F^{(1)}_2(D^*)\right) \right], \nonumber \\
\chi_{pt,4}^{(3)} &=& - \frac{N|d_{10}|^2}{\hbar \epsilon_0} \left[ \nu F^{(2)}_2(B)\right.- \nonumber \\
&-& \eta \left(AF^{(1)}_2(B)-CF^{(1)}_2(D)\right)+\nonumber \\
&+& \eta'S_c^2 \left(AF^{(1)}_2(B)-C^*F^{(1)}_2(D^*)\right)- \nonumber \\
&+& \eta'\Omega_c^+\Omega_c^-A\left(F^{(1)}_1(B)+F^{(1)}_3(B)\right) - \nonumber \\
&-& \left. \eta'\Omega_c^+\Omega_c^-C^*\left(F^{(1)}_1(D^*)+F^{(1)}_3(D^*)\right) \right], \nonumber \\
\chi_{pp,0}^{(3)} &=& \frac{N|d_{10}|^2}{\hbar \epsilon_0} \frac{\nu'}{\Delta_{20}^*} \left[ S_c^2 F^{(2)}_0(B)+2\Omega_c^+\Omega_c^-F^{(2)}_1(B)\right], \nonumber \\
\chi_{pp,2}^{(3)} &=& \frac{N|d_{10}|^2}{\hbar \epsilon_0} \frac{\nu'}{\Delta_{20}^*} \left[ S_c^2 F^{(2)}_1(B)+\Omega_c^+\Omega_c^-\left(F^{(2)}_0(B)+F^{(2)}_2(B)\right)\right], \nonumber \\
\chi_{pp,4}^{(3)} &=& \frac{N|d_{10}|^2}{\hbar \epsilon_0} \frac{\nu'}{\Delta_{20}^*} \left[ S_c^2 F^{(2)}_2(B)+\Omega_c^+\Omega_c^-\left(F^{(2)}_1(B)+F^{(2)}_3(B)\right)\right], \nonumber
\end{eqnarray}
As shown above, only these coefficients are present in the pulse propagation equations. Note that the corresponding formulae given in Ref. \cite{myScripta} included some misprints.

\begin{center}{APPENDIX B} \end{center}
\setcounter{equation}{0}
\renewcommand{\theequation}{B.\arabic{equation}}
To evaluate the population redistribution we proceed as follows.
We first use equations 2-4 of the stationary version of Eqs. (1)
to express $\sigma_{00}$ through $\sigma_{j0}$ and $\sigma_{jj}$:
\begin{eqnarray}
\sigma_{00}&=&\frac{2}{\gamma_{11}} \Im(\Omega_p \sigma_{10})+\xi(\sigma_{11}-\sigma_{33}), \nonumber \\
&=&\frac{2}{\gamma_{22}} \Im(\Omega_c \sigma_{20}) -\xi(\sigma_{11}+\sigma_{33}), \nonumber \\
&=&\frac{2}{\gamma_{33}} \Im(\Omega_t \sigma_{30}) +2\xi \sigma_{33}, \nonumber
\end{eqnarray}
where $\xi = \frac{\gamma_{ij}}{\gamma_{kk}}$, $i,j,k = 1,2,3$. Note that here we do not assume any of the populations negligible. Next we express $\sigma_{j0}$ through the diagonal elements of $\sigma$ in the lowest-order approximation with respect to the probe and trigger:
\begin{eqnarray}
\sigma_{10}&=&\frac{\Omega_p^*(\sigma_{11}-\sigma_{00})}{\Delta_{10}^*-\frac{|\Omega_{c}|^2}{\Delta_{12}^*}}- \frac{\Omega_p^* |\Omega_c|^2(\sigma_{22}-\sigma_{00})}{D^*(\Delta_{10}^*\Delta_{12}^*-|\Omega_c|^2)}, \nonumber \\
\sigma_{20}&=&\frac{\Omega_c^*}{D} \left[ (\sigma_{22} - \sigma_{00}) + \frac{|\Omega_p|^2(\sigma_{11} - \sigma_{00})}{\Delta_{10}\Delta_{12} - |\Omega_c|^2} - \frac{|\Omega_p|^2|\Omega_c|^2(\sigma_{22} - \sigma_{00})}{\Delta_{12}D(\Delta_{10}\Delta_{12}-|\Omega_c|^2)} + \right.\nonumber \\
&+& \left. \frac{|\Omega_t|^2(\sigma_{33} - \sigma_{00})}{\Delta_{30}\Delta_{32} - |\Omega_c|^2} - \frac{|\Omega_t|^2|\Omega_c|^2(\sigma_{22} - \sigma_{00})}{\Delta_{32}D(\Delta_{30}\Delta_{32}-|\Omega_c|^2)}\right], \nonumber \\
\sigma_{30}&=&\frac{\Omega_t^*(\sigma_{33}-\sigma_{00})}{\Delta_{30}^*-\frac{|\Omega_{c}|^2}{\Delta_{32}^*}}- \frac{\Omega_t^* |\Omega_c|^2(\sigma_{22}-\sigma_{00})}{D^*(\Delta_{30}^*\Delta_{32}^*-|\Omega_c|^2)}, \nonumber
\end{eqnarray}
where $D = \Delta_{20}^*+\frac{|\Omega_p|^2}{\Delta_{12}}+\frac{|\Omega_t|^2}{\Delta_{32}}$.
Combining the two latter steps we obtain a set of linear equations for the diagonal elements:
\begin{eqnarray}
\sigma_{00} &=& \alpha (\sigma_{11}-\sigma_{00}) - \alpha_1 (\sigma_{22} - \sigma_{00}) + \xi(\sigma_{11}-\sigma_{33}), \nonumber \\
\sigma_{00} &=& \beta (\sigma_{33}-\sigma_{00}) - \beta_1 (\sigma_{22} - \sigma_{00}) + 2\xi\sigma_{33}, \label{UR} \\
\sigma_{00} &=& \gamma (\sigma_{11}-\sigma_{00})+\delta(\sigma_{33}-\sigma_{00})+\epsilon(\sigma_{22}-\sigma_{00}), \nonumber \\
\sigma_{00}&+&\sigma_{11}+\sigma_{22}+\sigma_{33} = 1, \nonumber
\end{eqnarray}
where:
\begin{eqnarray}
\alpha&=& \frac{2}{\gamma_{11}}|\Omega_p|^2\Im\left( \frac{1}{\Delta_{10}^*-\frac{|\Omega_{c}|^2}{\Delta_{12}^*}}\right),\nonumber\\
\alpha_1 &=& \frac{2}{\gamma_{11}}|\Omega_p|^2 |\Omega_c|^2 \Im\left( \frac{1}{D^*(\Delta_{10}^*\Delta_{12}^*-|\Omega_{c}|^2)}\right),\nonumber\\
\beta&=&\frac{2}{\gamma_{33}}|\Omega_t|^2\Im\left( \frac{1}{\Delta_{30}^*-\frac{|\Omega_{c}|^2}{\Delta_{32}^*}}\right),\nonumber\\
\beta_1&=&\frac{2}{\gamma_{33}}|\Omega_t|^2|\Omega_c|^2\Im\left( \frac{1}{D^*(\Delta_{30}^*\Delta_{32}^*-|\Omega_{c}|^2)}\right),\nonumber\\
\gamma&=&\frac{2}{\gamma_{22}}|\Omega_c|^2|\Omega_p|^2\Im\left( \frac{1}{D(\Delta_{10}\Delta_{12}-|\Omega_{c}|^2)}\right),\label{diag2}\\
\delta&=&\frac{2}{\gamma_{22}}|\Omega_c|^2|\Omega_t|^2\Im\left( \frac{1}{D(\Delta_{30}\Delta_{32}-|\Omega_{c}|^2)}\right),\nonumber\\
\epsilon&=&\frac{2}{\gamma_{22}}|\Omega_{c}|^2 \left[\Im \left(\frac{1}{D}\right)-|\Omega_c|^2|\Omega_p|^2\Im\left(\frac{1}{\Delta_{12}D^2(\Delta_{10}\Delta_{12}-|\Omega_c|^2)}\right)- \right.\nonumber \\
&+& \left. |\Omega_c|^2|\Omega_t|^2\Im\left(\frac{1}{\Delta_{32}D^2(\Delta_{30}\Delta_{32}-|\Omega_c|^2)}\right)\right].\nonumber
\end{eqnarray}
The solutions of Eqs. (\ref{UR}) are
\begin{eqnarray}
\sigma_{11}&=& \frac{1}{M}\left\{ (1+\alpha)(\beta_1 \delta + \beta \epsilon) + (3+2\alpha + \beta)\epsilon \xi + \beta_1(-\alpha+\delta+\gamma)\xi+ \right.\nonumber \\ &+&\left.\alpha_1\left[-\delta+3\xi+2(\delta+\gamma)\xi+\beta(1+\gamma+\xi)\right]\right\},\nonumber\\
\sigma_{22}&=&\frac{1}{M}\left\{ (-1+\alpha_1)\beta \gamma +\left[(-1+\beta_1)\delta +\beta(2-\alpha_1+\epsilon)+\gamma(-3+2\alpha_1+\beta_1)\right]\xi + \right. \nonumber \\
&+& 2(3-\alpha_1-\beta_1+\beta+\delta+\gamma+\epsilon)\xi^2+ \nonumber \\
&+& \left. \alpha\left[(-1+\beta_1)\delta+(3-\beta_1)\xi+2\xi(\delta+\epsilon+\xi)+\beta(1+\epsilon+2\xi)\right]\right\},\label{diag1}\\
\sigma_{33}&=&\frac{1}{M} \left\{ -\beta_1 \gamma +\alpha_1(1+\beta)(\gamma-\xi)+\left[(1+\beta)\epsilon+\beta_1(2+\delta+\gamma)\right]\xi+\right. \nonumber \\
&+& \left.\alpha\left[(1+\beta)\epsilon+\beta_1(1+\delta+\epsilon)\right] \right\},\nonumber
\end{eqnarray}
while the value of $\sigma_{00}$ follows from the probability conservation. The denominator $M$ reads:
\begin{eqnarray}
M &=& \beta_1(\delta-\gamma)+\beta(\epsilon-\gamma+4\alpha_1\gamma)+\alpha_1(\beta+\gamma-\delta)+ \nonumber \\
 &+& \left[2\beta-\delta+4\epsilon+4\beta\epsilon-3\gamma+2\alpha_1(1-\beta+3\gamma+\delta)+\beta_1(2+4\gamma+4\delta)\right]\xi + \nonumber \\
 &+& 2(3-2\alpha_1-2\beta_1+\beta+\gamma+\delta+2\epsilon)\xi^2+ \nonumber \\
 &+& \alpha\left[\beta-\delta+\epsilon+4\beta\epsilon+\beta_1(1+4\delta-2\xi)+3\xi +2\xi(\beta+\delta+3\epsilon+\xi)\right]. \nonumber
 \end{eqnarray}
Using the above equations with constant initial values of probe and trigger fields and a given value of the control field we are able to estimate the population distribution.

\newpage

\section*{References}
\begin{thebibliography}{100}
\bibitem{fleischhauer}
M. Fleischhauer, A. Imamoglu and J. P. Marangos, Rev. Mod. Phys. {\bf 77}, 633 (2005).
\bibitem{my04}
A. Raczy\'nski, J. Zaremba and S. Zieli\'nska - Kaniasty, Phys. Rev. A \textbf{69} 043801 (2004).
\bibitem{paspalakis}
E. Paspalakis and P. L. Knight, J. Opt. B: Quantum Semiclassical Opt. {\bf 4}, S372 (2002).
\bibitem{mazets}
I. E. Mazets, Phys. Rev. A {\bf 71}, 023806 (2005).
\bibitem{my07}
A. Raczy\'nski, J. Zaremba and S. Zieli\'nska-Kaniasty, Phys. Rev. A {\bf 75} 013810 (2007).
\bibitem{kang}
H. Kang and Y. Zhu, Phys. Rev. Lett. \textbf{91} 093601 (2003).
\bibitem{wilson}
C. Goren, A. D. Wilson - Gordon, M. Rosenbluh and H. Friedmann, Phys. Rev. A \textbf{69} 053818 (2004).
\bibitem{joshi}
A. Joshi and M. Xiao, Phys. Lett. A {\bf 317}, 370 (2003).
\bibitem{ottaviani}
C. Ottaviani, D. Vitali, M. Artoni, F. Cataliotti and P. Tombesi,
Phys. Rev. Lett. A {\bf 19}, 197902 (2003).
\bibitem{rebic}
S. Rebi\'c, D. Vitali, C. Ottaviani, P. Tombesi, M. Artoni, F. Cataliotti and R. Corbalan, Phys. Rev. A {\bf 70}, 032317 (2004).
\bibitem{petr}
D. Petrosyan and Y. P. Malakyan, Phys. Rev. A, \textbf{70}, 023822 (2004).
\bibitem{joshi2}
A. Joshi and M. Xiao, Phys. Rev. A, \textbf{72}, 062319 (2005).
\bibitem{yang}
X. Yang, S. Li, C. Zhang and H. Wang, J. Opt. Soc. Am. B \textbf{26}, 1423 (2009).
\bibitem{kou}
J. Kou, R. G. Wan, Z. H. Kang, H. H. Wang, L. Jiang, X. J. Zhang, Y. Jiang and J. Y. Gao, J. Opt. Soc. Am. B \textbf{27}, 2035 (2010).
\bibitem{lo}
H.-Y. Lo, P.-C. Su and Y.-F. Chen, Phys. Rev. A, \textbf{81},
053829 (2010).
\bibitem{wang}
Z.-B. Wang, K.-P. Marzlin and B. C. Sanders, Phys. Rev. Lett. \textbf{97}, 063901 (2006).
\bibitem{liu}
C. Liu, Z. Dutton, C. H. Behroozi and L. V. Hau, Nature {\bf 409},
490 (2001).
\bibitem{matsko}
A. B. Matsko, O. Kocharovskaya, Y. Rostovtsev, G. L. Welch, A. S.
Zibrov and M. O. Scully, Adv. At. Mol. Opt. Phys. {\bf 46}, 191,
(2001).
\bibitem{andre} A. Andr\'e and M. D. Lukin, Phys. Rev.
Lett. \textbf{89}, 143602 (2002).
\bibitem{my09}
J.-H. Wu, A. Raczy\'nski, J. Zaremba, S. Zieli\'nska-Kaniasty, M.
Artoni and G. C. La Rocca, J. Mod. Opt. {\bf 56}, 768 (2009); and
references therein.
\bibitem{myNowa}
A. Raczy\'nski, J. Zaremba, S. Zieli\'nska-Kaniasty, M. Artoni and
G. C. La Rocca,  J. Mod. Opt. \textbf{56}, 2348 (2009).
\bibitem{artoni2}
M. Artoni and G. C. La Rocca, Phys. Rev. Lett. {\bf 96}, 073905
(2006).
\bibitem{bajcsy} M. Bajcsy, A. S. Zibrov and M. D. Lukin,
Nature \textbf{426}, 638 (2003).
\bibitem{newchen}
A. Andr\'e and M. D. Lukin, Phys. Rev. Lett. \textbf{94},
063902(2005); I. Friedler, G. Kurizki and D. Petrosyan, Phys. Rev.
A, \textbf{71}, 023803 (2005); Y.-F. Chen, C.-Y. Wang, S.-H. Wang
and I. A. Yu, Phys. Rev. Lett. \textbf{96}, 043603 (2006).
\bibitem{wucav}
J.-H. Wu, M. Artoni, G. C. La Rocca, Phys. Rev. Lett. {\bf 103},
133601 (2009).
\bibitem{myScripta}
K. S\l owik, A. Raczy\'nski, J. Zaremba, S. Zieli\'nska - Kaniasty, M. Artoni and G. C. La Rocca, Phys. Scr. T \textbf{143}, 014022 (2011).
\end{thebibliography}
\end{document}